\documentclass{elsart}
\usepackage{hyperref}

\usepackage{amssymb}
\usepackage[dvips]{graphicx}
\usepackage{dcolumn}
\usepackage{bm}
\usepackage{amsmath}
\begin{document}

\begin{frontmatter}


\title{Estimating the mass of vortices in the cuprate superconductors}

\author[Harvard,Frankfurt]{Lorenz Bartosch},
\author[UCSB]{Leon Balents}, and
\author[Harvard]{Subir Sachdev}

\address[Harvard]{Department of Physics, Harvard University, Cambridge MA
02138}

\address[Frankfurt]{Institut f\"ur Theoretische Physik, Universit\"at
Frankfurt, Postfach 111932, 60054 Frankfurt, Germany}

\address[UCSB]{Department of Physics, University of California, Santa
Barbara, CA 93106-4030}


\begin{abstract}
We explore the experimental implications of a recent theory of the
quantum dynamics of vortices in superfluids proximate to Mott
insulators. The theory predicts modulations in the local density
of states in the regions over which the vortices execute their
quantum zero point motion. We use the spatial extent of such
modulations in scanning tunnelling microscopy measurements
(Hoffman {\em et al\/}, Science {\bf 295}, 466 (2002)) on the
vortex lattice of Bi$_2$Sr$_2$CaCu$_2$O$_{8+\delta}$ to estimate
the inertial mass of a point vortex. We discuss other, more
direct, experimental signatures of the vortex dynamics.
\end{abstract}


\end{frontmatter}

\section{Introduction}

It is now widely accepted that superconductivity in the cuprates is
described, as in the standard Bardeen-Cooper-Schrieffer (BCS)
theory, by the condensation of charge $-2e$ Cooper pairs of
electrons. However, it has also been apparent that vortices in the
superconducting state are not particularly well described by BCS
theory. While elementary vortices do carry the BCS flux quantum of
$hc/2e$, the local electronic density of states in the vortex core,
as measured by scanning tunnelling microscopy (STM) experiments, has
not been explained naturally in the BCS framework. Central to our
considerations here are the remarkable STM measurements of Hoffman
{\em et al.} \cite{hoffman} (see also Refs.~\cite{fischer,pan}) who
observed modulations in the local density of states (LDOS) with a
period of approximately 4 lattice spacings in the vicinity of each
vortex core of a vortex lattice in
Bi$_2$Sr$_2$CaCu$_2$O$_{8+\delta}$.

This paper shall present some of the physical implications of a
recent theory of two-dimensional superfluids in the vicinity of a
quantum phase transition to a Mott insulator \cite{bbbss1,bbbss2}
(see also Ref.~\cite{zlatko}). By `Mott insulator' we mean here an
incompressible state which is pinned to the underlying crystal
lattice, with an energy gap to charged excitations. In the Mott
insulator, the average number of electrons per unit cell of the
crystal lattice, $n_{MI}$, must be a rational number. If the Mott
insulator is not `fractionalized' and if $n_{MI}$ is not an even
integer, then the Mott insulator must also spontaneously break the
space group symmetry of the crystal lattice so that the unit cell
of the Mott insulator has an even integer number of electrons.
There is evidence that the hole-doped cuprates are proximate to a
Mott insulator with $n_{MI} = 7/8$ \cite{jtran}, and such an
assumption will form the basis of our analysis of the STM
experiments on Bi$_2$Sr$_2$CaCu$_2$O$_{8+\delta}$. The electron
number density in the superfluid state, $n_S$, need not equal
$n_{MI}$ and will be assumed to take arbitrary real values, but
not too far from $n_{MI}$.

\begin{figure}
\centering
\includegraphics[width=4in]{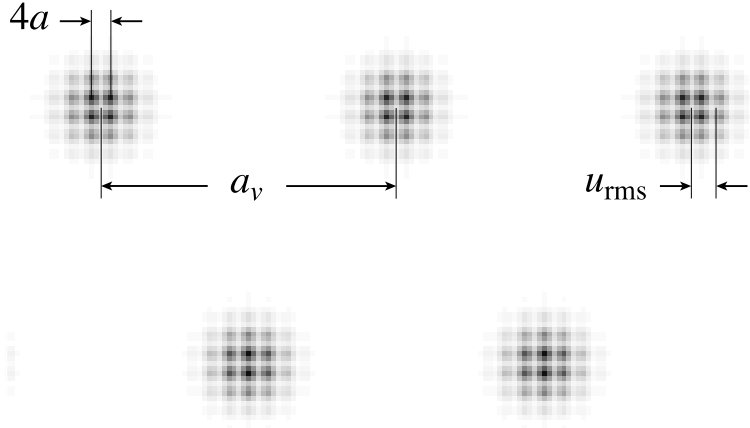}
\caption{Schematic of the modulations in the LDOS of a vortex
lattice. The short distance modulations in each vortex halo are
determined by the orientation of the vortex in flavor space, as
discussed in Ref.~\cite{bbbss1}. The envelope of these modulations
is $|\Psi ({\bf r}_j) |^2$ where $\Psi$ is the wavefunction of the
vortices, and its characteristics are computed in the present
paper.} \label{figfluxlattice}
\end{figure}

A key ingredient in our analysis will be the result that the
superfluid carries a subtle quantum order, which is distinct from
Landau-Ginzburg order of a Cooper pair condensate. In two
dimensions, vortices are point-like excitations, and are therefore
bona fide quasiparticle excitations of the superfluid. The quantum
order is reflected in the wavefunction needed to describe the
motion of the vortex quasiparticle. For $n_{MI}$ not an even
integer, the low energy vortices appear in multiple degenerate
flavors, and the space group symmetry of the underlying lattice is
realized in a projective unitary representation that acts on this
flavor space. Whenever a vortex is pinned (either individually due
to impurities, or collectively in a vortex lattice), the space
group symmetry is locally broken, and hence the vortex necessarily
chooses a preferred orientation in its flavor space. As shown in
Ref.~\cite{bbbss1}, this implies the presence of modulations in
the LDOS in the spatial region over which the vortex executes its
quantum zero point motion \cite{vbs}. The short-distance structure
and period of the modulations is determined by that of the Mott
insulator at density $n_{MI}$, while its long-distance envelope is
a measure of the amplitude of the vortex wavefunction (see
Fig.~\ref{figfluxlattice}).
Consequently, the size of the region where the modulations are
present is determined by the inertial mass of the vortex. Here we
will show how these ideas can be made quantitatively precise, and
use current experiments to obtain an estimate of the vortex mass,
$m_v$. There have been a number of theoretical discussions of
$m_v$ using BCS theory
\cite{volovik,kopnin,blatter,duan,tvr,Coffey98,han},
and they lead to the order of magnitude estimate $m_v \sim m_e
(k_F \xi)^2$, where $m_e$ is the electron mass, $k_F$ is the Fermi
wavevector, and $\xi$ is the BCS coherence length.

\section{Vortex equations of motion}

We begin with a simple model computation of the vortex dynamics.
Consider a system of point vortices moving in a plane at positions
${\bf r}_j$, where $j$ is a label identifying the vortices. We do
not explicitly identify the orientation of each vortex in flavor
space, because we are interested here only in the long-distance
envelope of the LDOS modulations; the flavor orientation does not
affect the interactions between well-separated vortices, and so
plays no role in determining the wavefunction of the vortex
lattice. In a Galilean-invariant superfluid, the vortices move
under the influence of the Magnus force
\begin{equation}
m_v \frac{d^2 {\bf r}_j}{d t^2} = \frac{h n_S}{2 a^2} \left( {\bf
v}_s ({\bf r}_j ) - \frac{d {\bf r}_j}{dt} \right) \times \hat{\bf
z}, \label{e1}
\end{equation}
where $t$ is time, $h=2\pi \hbar$ is Planck's constant, ${\bf v}_s
({\bf r})$ is the superfluid velocity at the position ${\bf r}$,
and $n_S/a^2$ is the electron number density per unit area ($a^2$
is the area of a unit cell of the underlying lattice). One point
of view is that the force in Eq.~(\ref{e1}) is that obtained from
classical fluid mechanics after imposing the quantization of
circulation of a vortex. However, Refs.~\cite{haldane} emphasized
the robust topological nature of the Magnus force and its
connection to Berry phases, and noted that it applied not only to
superfluids of bosons, but quite generally to superconductors of
paired electrons. Here, we need the modification of Eq.~(\ref{e1})
by the periodic crystal potential and the proximate Mott
insulator. This was implicit in the results of Ref.~\cite{bbbss1},
and we present it in more physical terms. It is useful to first
rewrite Eq.~(\ref{e1}) as
\begin{equation}
m_v \frac{d^2 {\bf r}_j}{d t^2} = {\bf F}_E (j) + {\bf F}_B (j),
\label{e2}
\end{equation}
where ${\bf F}_E$ is the first term proportional to ${\bf v}_s$
and ${\bf F}_B$ is the second term. Our notation here is
suggestive of a dual formulation of the theory in which the
vortices appear as `charges', and these forces are identified as
the `electrical' and `magnetic' components. In the Galilean
invariant superfluid, the values of ${\bf F}_E$ and ${\bf F}_B$
are tied to each other by a Galilean transformation. However, with
a periodic crystal potential, this constraint no longer applies,
and their values renormalize differently as we now discuss.

The influence of the crystal potential on ${\bf F}_E$ is simple,
and replaces the number density of electrons, $n_S$, by the
superfluid density. Determining ${\bf v}_s ({\bf r}_j )$ as a sum
of contributions from the other vortices, we obtain
\cite{background}
\begin{equation}
{\bf F}_E (j) = 2 \pi \rho_s \sum_{k\neq j} \frac{{\bf r}_j - {\bf
r}_k}{|{\bf r}_j - {\bf r}_k |^2}, \label{e3}
\end{equation}
where $\rho_s$ is the superfluid stiffness (in units of energy).
It is related to the London penetration depth, $\lambda$, by
\begin{equation}
\rho_s = \frac{\hbar^2 c^2 d}{16 \pi e^2 \lambda^2},
\end{equation}
where $d$ is the interlayer spacing.

The modification of ${\bf F}_B$ is more subtle. This term states
that the vortices are `charges' moving in a `magnetic' field with
$n_S/2$ `flux' quanta per unit cell of the periodic crystal
potential. In other words, the vortex wavefunction is obtained by
diagonalizing the Hofstadter Hamiltonian which describes motion of
a charged particle in the presence of a magnetic field and a
periodic potential. As argued in Ref.~\cite{bbbss1}, it is useful
to examine this motion in terms of the deviation from the rational
`flux' $n_{MI}/2=p/q$ ($p$, $q$ are relatively prime integers)
associated with the proximate Mott insulator. The low energy
states of the rational flux Hofstadter Hamiltonian have a $q$-fold
degeneracy, and this constitutes the vortex flavor space noted
earlier \cite{sflux}. However, these vortex states describe
particle motion in zero `magnetic' field, and only the deficit
$(n_{S} - n_{MI})/2$ acts as a `magnetic flux'. This result is
contained in the action in Eq.~(2.46) of Ref.~\cite{bbbss1}, which
shows that the dual gauge flux fluctuates about an average flux
determined by $(n_{S} - n_{MI})$. The action in Ref.~\cite{bbbss1}
has a `relativistic' form appropriate to a system with equal
numbers of vortices and anti-vortices. Here, we are interested in
a system of vortices induced by an applied magnetic field, and can
neglect anti-vortices; so we should work with the corresponding
`non-relativistic' version of Eq.~(2.46) of Ref.~\cite{bbbss1}. In
its first-quantized version, this `non-relativistic' action for
the vortices leads to the `Lorentz' force in Eq.~(\ref{e2}) given
by
\begin{equation}
{\bf F}_B (j) =  -\frac{h (n_S  -n_{MI})}{2 a^2}
\frac{d {\bf r}_j}{dt} \times \hat{\bf z}, \label{e4}
\end{equation}
If the density of the superfluid equals the commensurate density of
the Mott insulator, then ${\bf F}_B = 0$; however, ${\bf F}_E$ remains
non-zero because we can still have $\rho_s \neq 0$ in the superfluid.
These distinct behaviors of ${\bf F}_{E,B}$ constitute a key
difference from Galilean-invariant superfluids. In experimental
studies of vortex motion in superconductors \cite{ong}, a force of the
form of Eq.~(\ref{e4}) is usually quoted in terms of a `Hall drag'
co-efficient per unit length of the vortex line, $\alpha$;
Eq.~(\ref{e4}) implies
\begin{equation}
\alpha = - \frac{h (n_S - n_{MI})}{2a^2 d}. \label{e5}
\end{equation}
Thus the periodic potential has significantly reduced the magnitude of
$\alpha$ from the value nominally expected \cite{kopkra} by
subtracting out the density of the Mott insulator. A smaller than
expected $|\alpha|$ is indeed observed in the cuprates \cite{ong}.  It
is worth emphasizing that ${\bf F}_B$ (but not ${\bf F}_E$) is an {\sl
  intrinsic} property of a single vortex.  Moreover, we expect that,
taken together, the relation Eq.~(\ref{e5}) and the flavor degeneracy
$q$ are robust ``universal'' measures of the quantum order of a clean
superconductor, independent of details of the band structure, etc.

\section{Vortex lattice normal modes}

\begin{figure}
\centering
\includegraphics[width=4in]{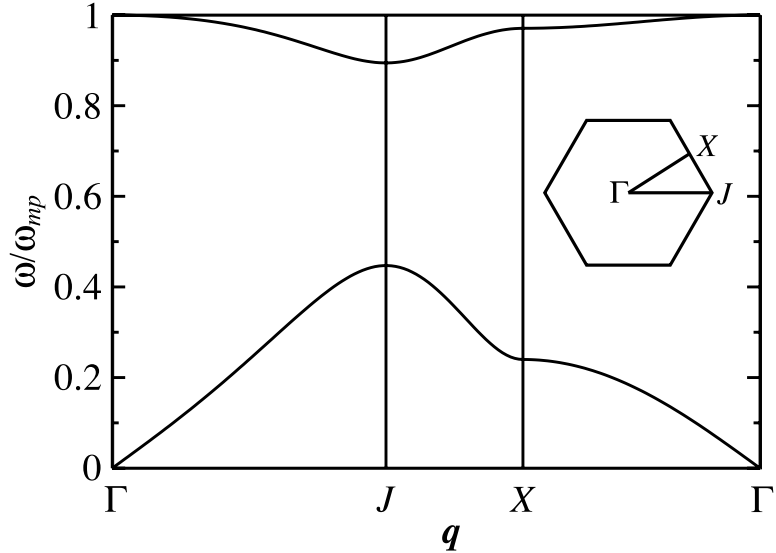}
\caption{Dispersion of the `magnetophonon' modes of the vortex
lattice (here we take $\omega_c=0.5 \omega_p$).} \label{figdispersion}
\end{figure}

We solved Eqs.~(\ref{e2}-\ref{e4}) in the harmonic approximation
in deviations from a perfect triangular lattice of vortices at
positions ${\bf R}_j$. This leads to the vortex `magnetophonon'
modes shown in Fig~\ref{figdispersion}. The computation of these
modes is a generalization of other vortex oscillation modes
discussed previously in superconductors \cite{degennes,fetter},
rotating superfluids \cite{fetter1,tkachenko,baym}, and, in a dual
picture of `charges', also to oscillations of electronic Wigner
crystals in a magnetic field \cite{fukuyama}.
Quantizing these modes, we determine the mean square displacement of
each vortex due to the quantum zero point motion of the vortex
lattice, which we denote $u_{\text{rms}}^2 = \langle |{\bf r}_j -
{\bf R}_j |^2 \rangle/2$. We found an excellent fit of this solution
to the interpolation formula
\begin{equation}
m_v = \frac{0.03627 a_v^2 \hbar^2}{\rho_s u_{\text{rms}}^4} F(x),
\label{e6}
\end{equation}
where $a_v$ is the separation between nearest neighbor vortices,
$x \equiv |\alpha| d u_{\text{rms}}^2/\hbar$, and
\begin{equation}
F(x) \approx 1 - 0.4099 x^2 ; \label{e7}
\end{equation}
Eq.~(\ref{e7}) holds only as long as the r.h.s. is positive, while
$F(x)=0$ for larger $x$ (we will see below and in
Fig.~\ref{graphF} that this apparent upper bound on $x$ is relaxed
once we allow for viscous damping). Similarly, for the frequency
$\omega_{mp}$ in Fig.~\ref{figdispersion}, we obtain $\omega_{mp}
= \sqrt{\omega_p^2 + \omega_c^2}$ with
\begin{equation}
\omega_p = \frac{35.45 \rho_s u_{\text{rms}}^2}{\hbar a_v^2
[F(x)]^{1/2}}~~;~~\omega_c = \frac{27.57 \rho_s u_{\text{rms}}^2 x
F(x)}{\hbar a_v^2}. \label{e8}
\end{equation}
For the experiments of Ref.~\cite{hoffman} we estimate $\rho_s =
12$ meV \cite{weber}, $u_{\text{rms}} = 20$ \AA\ \cite{urms}, $a =
3.83$ \AA, and $a_v=240$ \AA. The overall scale for $m_v$ is
determined by setting $n_S=n_{MI}$ so that $x=0$ and $F(x)=1$.
This yields $m_v \approx 8 m_e$ and $\hbar \omega_p \approx 3$ meV
(or $\nu_p \approx 0.7$ THz). For a more accurate determination,
we need $n_S$, for which there is considerable uncertainty {\em
e.g.\/} for $|n_S - n_{MI}| = 0.015$, we find $x=1.29$, $m_v
\approx 3 m_e$ and $\hbar \omega_p \approx 5$~meV.

\section{Limitations}

We now consider the influence of a variety of effects which have
been neglected in this simple computation:

\subsection{Viscous drag}

\begin{figure}
\centering
\includegraphics[width=4in]{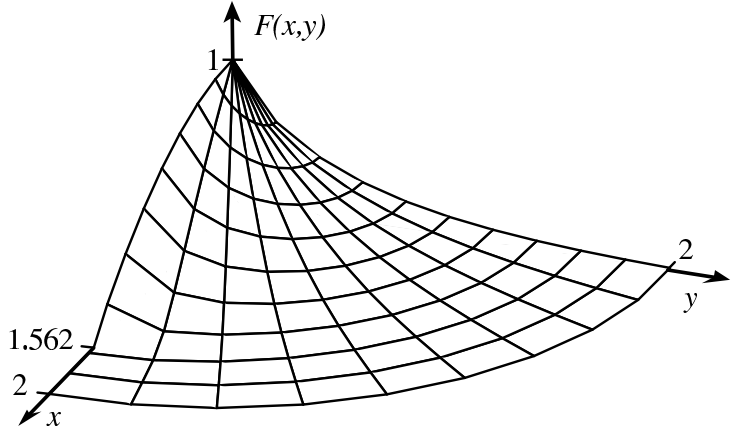}
\caption{Plot of the function $F(x,y)$ which replaces $F(x)$ in
Eqs. (\ref{e6},\ref{e7},\ref{e8}) upon including viscous drag,
$\eta$ ($y \equiv \eta d u_{\rm rms}^2 /\hbar$). The argument $x$
measures the Hall drag $\alpha$ ($x \equiv |\alpha| d u_{\rm
rms}^2 /\hbar$) and the Hall angle is determined by $|\tan
\theta_H | = x/y$.} \label{graphF}
\end{figure}

It is conventional in models of vortex dynamics at low frequencies
\cite{bardeen} to include a dissipative viscous drag term in the
equations of motion, contributing an additional force
\begin{equation}
{\bf F}_D (j) = - \eta d \frac{d {\bf r}_j}{d t}, \label{e9}
\end{equation}
to the r.h.s. of Eq.~(\ref{e2}). There are no reliable theoretical
estimates for the viscous drag co-efficient, $\eta$, for the
cuprates. However, we can obtain estimates of its value from
measurements of the Hall angle, $\theta_H$, which is given by
\cite{bardeen,ong2} $\tan \theta_H = \alpha/\eta$. Harris {\em et
al.} \cite{ong2} observed a dramatic increase in the value $|\tan
\theta_H|$, to the value 0.85, at low $T$ in ``60 K''
YBa$_2$Cu$_3$O$_{6+y}$ crystals, suggesting a small $\eta$, and weak
dissipation in vortex motion. For our purposes, we need the value of
$\eta$ for frequencies of order $\omega_p$, and not just in the d.c.
limit. The very na\"ive expectation that $\eta(\omega)$ behaves like
the quasiparticle microwave conductivity would suggest it decreases
rapidly beyond a few tens of GHz, well below $\omega_p$
\cite{turner03}. Lacking solid information, we will be satisfied
with an estimate of the influence of viscous drag obtained by
neglecting the frequency dependence of $\eta$ (a probable
overestimate of its influence). The resulting corrections to
Eqs.~(\ref{e6}-\ref{e8}) are easily obtained (as in
Ref.~\cite{blatter}), and can be represented by the replacement
\begin{equation}
F(x) \rightarrow F(x,y),~\mbox{where}~y \equiv \frac{\eta d
u_{\text{rms}}^2}{\hbar}
\end{equation}
and
\begin{equation}
F(x,0)=F(x).
\end{equation}
The sketch of the function $F(x,y)$ is in Fig.~\ref{graphF}; as long
as $y>0$, we have $F(x,y)>0$.
As expected, the viscous damping decreases the estimate of the mass,
and this decrease is exponential for large $y$, {\em e.g.\/} at
$x=0$ we have the interpolation formula
\begin{equation}
F(0,y)
\approx (1 + 0.41 y + 2.69y^2) e^{-3.43 y}.
\end{equation}

\subsection{Meissner screening}

The interaction in Eq.~(\ref{e3}) is screened at long distances by
the supercurrents, and the intervortex coupling becomes
exponentially small. This does have an important influence at small
momenta in that the shear mode of the vortex lattice disperses as
\cite{degennes,fetter} $\sim k^2$. However, as long as $a_v \ll
\lambda$, there will not be a significant influence on
$u_{\text{rms}}$ or $\omega_{p}$.

\subsection{Retardation}

The interaction Eq.~(\ref{e3}) is assumed to be instantaneous; in
reality it is retarded by the propagation of the charged plasmon
mode of the superfluid. We have estimated the corrections due to
this mode in a model of superfluid layers coupled by the long-range
Coulomb interaction. Briefly, we compare the energy per unit area of
a `phase fluctuation' at the wavevector of the vortex lattice
Brillouin zone boundary ($\sim \rho_s /a_v^2$) with its
electrostatic energy ($\sim \hbar^2 \omega_p^2 /(e^2 d)$); this
shows that such corrections are of relative order $\sim
(\hbar^2/m_v)/(e^2 d)$. For the parameters above and $d=7.5$ \AA\
this is $\sim 0.009$, and hence quite small.

\subsection{Nodal quasiparticles}

We expect that nodal quasiparticles contribute to the viscous drag,
and so their contribution was already included in the experimentally
determined estimate of $\eta$ in ({\em i\/}). The quasiparticle
contribution to $m_v$ has been estimated \cite{volovik,kopnin} in a
BCS theory of a continuum of core levels, valid in the limit $k_F
\xi \gg 1$, and a $\sim 1/\sqrt{H}$ dependence ($H$ is the magnetic
field) was obtained. The applicability of such a model to the
cuprates is questionable, but nevertheless the estimates of $m_v$
above should provisionally be assumed to be specific to the $H$
field in the STM experiments.

\subsection{Disorder}

We have assumed here a triangular lattice of vortices. In reality,
STM experiments show significant deviations from such a structure,
presumably because of an appreciable random pinning potential. This
pinning potential will also modify the vortex oscillation
frequencies and its mean square displacement. Both pinning and
damping $\eta$ tend to reduce vortex motion.  For this reason, the
estimates of $m_v$ above in which these effects are neglected must
be regarded as upper bounds.

The above considerations make it clear that new experiments on
cleaner underdoped samples, along with a determination of the
spatial dependence of the hole density (to specify $\alpha$), are
necessary to obtain a more precise value for $m_v$; determining
the $H$ dependence of $m_v$ will enable confrontation with theory.

\section{Implications}

An important consequence of our theory is the emergence of
$\omega_p$ as a characteristic frequency of the vortex dynamics.
It would therefore be valuable to have an inelastic scattering
probe which can explore energy transfer on the scale of $\hbar
\omega_p$, and with momentum transfer on the scale of $h/a_v$,
possibly by neutron \cite{neutron} or X-ray scattering. A direct
theoretical consideration of magnetoconductivity in our picture
would have implications for far-infrared or THz spectroscopy,
allowing comparison to existing experiments \cite{Lihn96}; further
such experiments on more underdoped samples would also be of
interest.

Another possibility is that the zero point motion of the vortices
emerges in the spectrum of the LDOS measured by STM at an energy
of order $\hbar \omega_p$. We speculate that understanding the
`vortex core states' observed in STM studies \cite{fischer,pan}
will require accounting for the quantum zero point motion of the
vortices; it is intriguing that the measured energy of these
states is quite close to our estimates of $\hbar \omega_p$.

\section{Acknowledgements}

We thank J.~Brewer, E.~Demler, \O.~Fischer, M.~P.~A.~Fisher,
W.~Hardy, B.~Keimer, N.~P.~Ong, T.~Senthil, G.~Sawatzky, Z.~Te\v
sanovi\'c and especially J.~E.~Hoffman and J.~C.~Seamus Davis for
useful discussions. This research was supported by the NSF under
grants DMR-0457440 (L.~Balents), DMR-0098226 and DMR-0455678
(S.S.), the Packard Foundation (L.~Balents), the Deutsche
Forschungsgemeinschaft under grant BA 2263/1-1 (L.~Bartosch), and
the John Simon Guggenheim Memorial Foundation (S.S.).


\begin{thebibliography}{99}

\bibitem{hoffman} J.~E.~Hoffman, E.~W.~Hudson, K.~M.~Lang, V.~Madhavan,
S.~H.~Pan, H.~Eisaki, S.~Uchida, and J.~C.~Davis, Science {\bf
295}, 466 (2002).

\bibitem{fischer} B.~W.~Hoogenboom,
K.~Kadowaki, B.~Revaz, M.~Li, Ch.~Renner, and \O.~Fischer, Phys.
Rev. Lett. {\bf 87}, 267001 (2001); G.~Levy, M.~Kugler,
A.~A.~Manuel, \O.~Fischer, and M.~Li, preprint.

\bibitem{pan} S.~H.~Pan,
E.~W.~Hudson1, A.~K.~Gupta, K.-W.~Ng, H.~Eisaki, S.~Uchida, and
J.~C.~Davis, Phys. Rev. Lett. {\bf 85}, 1536 (2000).

\bibitem{bbbss1} L.~Balents, L.~Bartosch, A.~Burkov, S.~Sachdev,
and K.~Sengupta, Phys. Rev. B {\bf 71}, 144508 (2005).

\bibitem{bbbss2} L.~Balents, L.~Bartosch, A.~Burkov, S.~Sachdev,
and K.~Sengupta, Phys. Rev. B {\bf 71}, 144509 (2005).

\bibitem{zlatko} Z.~Te\v sanovi\'c, Phys. Rev. Lett. {\bf 93}, 217004 (2004);
M.~Franz, D.~E.~Sheehy, and Z.~Te\v sanovi\'c, Phys. Rev. Lett.
{\bf 88}, 257005 (2002).

\bibitem{jtran} J.~M.~Tranquada, B.~J.~Sternlieb, J.~D.~Axe, Y.~Nakamura,
and S.~Uchida, Nature {\bf 375}, 561 (1995); J.~M.~Tranquada,
H.~Woo, T.~G.~Perring, H.~Goka, G.~D.~Gu, G.~Xu, M.~Fujita, and
K.~Yamada, Nature {\bf 429}, 534 (2004).

\bibitem{vbs} The LDOS modulations are a signal of
valence-bond-solid (VBS) order that can be associated with the
vortex flavors \cite{bbbss1} (see also  C. Lannert,
M.~P.~A.~Fisher, and T.~Senthil, Phys. Rev. B {\bf 63}, 134510
(2001)). VBS order in vortex cores was predicted in K.~Park and
S.~Sachdev, Phys. Rev. B {\bf 64}, 184510 (2001).

\bibitem{volovik} G. E. Volovik, Pisma Zh. Eksp. Teor. Fiz. {\bf 65}
201 (1997) [JETP Lett. {\bf 65} 217 (1997)].

\bibitem{kopnin} N.~B.~Kopnin, {\em Theory of Nonequilibrium
Superconductivity}, Cambridge University Press, Cambridge (2001);
N.~B.~Kopnin and V.~M.~Vinokur, Phys. Rev. Lett. {\bf 81}, 3952
(1998).

\bibitem{blatter} G.~Blatter, V.~B.~Geshkenbein, and V.~M.~Vinokur,
Phys. Rev. Lett. {\bf 66}, 3297 (1991); G.~Blatter and B.~Ivlev,
Phys. Rev. Lett. {\bf 70}, 2621 (1993).

\bibitem{duan} J.-M.~Duan and A.~J.~Leggett, Phys. Rev. Lett. {\bf 68}, 1216
(1992).

\bibitem{tvr} D.~M.~Gaitonde and T.~V.~Ramakrishnan, Phys. Rev. B {\bf 56}, 11951
(1997).

\bibitem{Coffey98} M.~W.~Coffey, J.~Phys.~A {\bf 31}, 6103 (1998).

\bibitem{han} J.~H.~Han, J.~S.~Kim, M.~J.~Kim, and
P.~Ao, Phys. Rev. B {\bf 71}, 125108 (2005).


\bibitem{haldane} F.~D.~M.~Haldane and Y.-S.~Wu, Phys. Rev. Lett. {\bf 55}, 2887
(1985); P.~Ao and D.~J.~Thouless, Phys. Rev. Lett. {\bf 70}, 2158
(1993).

\bibitem{background} The long-range force
in Eq.~(\ref{e3}) requires that boundary conditions be chosen
carefully. In the dual charge analogy, we need a uniform
neutralizing background charge, and this contributes an additional
force $-2 \pi^2 \rho_s {\bf r}_j /A_v$ to Eq.~(\ref{e3}), where
$A_v = \sqrt{3} a_v^2/2$ is the area per vortex.

\bibitem{sflux} At $n_{MI}/2 = p/q$, the vortex degeneracy is $q$
 in the short-range pairing models
  of Ref.~\cite{bbbss2}.  In L.~Balents and
  S.~Sachdev, to appear, we show that a similar degeneracy can be
  understood in long-range pairing models of $d$-wave superconductors
  based upon the gauge theory of X.-G.~Wen and P.~A.~Lee, Phys.
  Rev. Lett. {\bf 76}, 503 (1996).  Any distinctions between these
  cases have no impact on the considerations of the present paper.

\bibitem{ong} J.~M.~Harris, N.~P.~Ong,
P.~Matl, R.~Gagnon, L.~Taillefer, T.~Kimura, and K.~Kitazawa,
Phys. Rev. B {\bf 51}, R12053 (1995).

\bibitem{kopkra} N.~B.~Kopnin and V.~E.~Kravtsov, Pis'ma Zh. Eksp.
Teor. Fiz. {\bf 23}, 631 (1976) [JETP Lett. {\bf 23}, 578 (1976)].

\bibitem{degennes} P.~G.~de Gennes and J.~Matricon, Rev. Mod.
Phys. {\bf 36}, 45 (1964).

\bibitem{fetter} A.~L.~Fetter, P.~C.~Hohenberg, and P.~Pincus,
Phys. Rev. {\bf 147}, 140 (1966); A.~L.~Fetter, Phys. Rev. {\bf
163}, 390 (1967).

\bibitem{fetter1} A.~L.~Fetter, Phys. Rev. {\bf
162}, 143 (1967).

\bibitem{tkachenko} V.~K.~Tkachenko, Zh. Eksp. Teor. Fiz. {\bf 49}, 1875 (1965)
[Sov. Phys. JETP {\bf 22}, 1282 (1966)].

\bibitem{baym} G.~Baym,
Phys. Rev. B {\bf 51}, 11697 (1995); Phys. Rev. Lett. {\bf 91},
110402 (2003).

\bibitem{fukuyama} H.~Fukuyama, Sol. State. Comm. {\bf 17}, 1323
(1975).

\bibitem{weber} M. Weber {\em et al.\/}, Phys. Rev. B {\bf 48}, 13022 (1993).

\bibitem{urms}  We estimate $u_{\text{rms}}$ from the published
data in Ref.~\cite{hoffman}. Also, using the correspondence between
the LDOS modulations and the vortex core states \cite{fischer}, we
obtain a similar number from the measurements of Ref.~\cite{pan}.

\bibitem{bardeen} J.~Bardeen and M.~J.~Stephen, Phys. Rev. {\bf
140}, A1197 (1965); P.~Nozi\`{e}res and W.~F.~Vinen, Philos. Mag.
{\bf 14}, 667 (1966).

\bibitem{ong2} J.~M.~Harris, Y.~F.~Yan, O.~K.~C.~Tsui, Y.~Matsuda,
and N.~P.~Ong, Phys. Rev. Lett. {\bf 73}, 1711 (1994).

\bibitem{turner03} P.~J.~Turner {\sl et al}, Phys. Rev. Lett. {\bf 90},
  237005 (2003).

\bibitem{neutron} B.~Keimer, W.~Y.~Shih, R.~W.~Erwin, J.~W.~Lynn,
F.~Dogan, and I.~A.~Aksay, Phys. Rev. Lett. {\bf 73}, 3459 (1994).

\bibitem{Lihn96} H.-T.~S.~Lihn, S.~Wu, H.~D.~Drew, S.~Kaplan, Qi~Li,
and D.~B.~Fenner, Phys. Rev. Lett. {\bf 76}, 3810 (1996) and
  references therein.


\end{thebibliography}
\end{document}